\documentstyle[epsf]{mn}

\begin{document}

\title[Large scale bias and the peak background split]
{Large scale bias and the peak background split}
\author[Ravi K. Sheth \& Giuseppe Tormen]
{Ravi K. Sheth$^1$ \& Giuseppe Tormen$^2$\\
$^1$ Max-Planck Institut f\"ur Astrophysik, 85740 Garching, Germany\\
$^2$ Dipartimento di Astronomia, 35122 Padova, Italy \\
\smallskip
Email: sheth@mpa-garching.mpg.de, tormen@pd.astro.it
}
\date{Submitted 20 December 1998}

\maketitle

\begin{abstract}
Dark matter haloes are biased tracers of the underlying 
dark matter distribution.  We use a simple model to provide a 
relation between the abundance of dark matter haloes and their 
spatial distribution on large scales.  Our model shows that 
knowledge of the unconditional mass function alone is sufficient 
to provide an accurate estimate of the large scale bias factor.  
Then we use the mass function measured in numerical simulations 
of SCDM, OCDM and $\Lambda$CDM to compute this bias.  Comparison 
with these simulations shows that this simple way of estimating 
the bias relation and its evolution is accurate for less massive 
haloes as well as massive ones.  In particular, we show that haloes 
which are less/more massive than typical $M_*$ haloes at the time 
they form are more/less strongly clustered than formulae based on 
the standard Press--Schechter mass function predict.  
\end{abstract}


\section{Introduction}
There has been considerable interest recently in developing 
models for the shape and evolution of the mass function of 
collapsed dark matter haloes (Press \& Schechter 1974; 
Bond et al. 1991; Lacey \& Cole 1993) as well as for the evolution 
of the spatial distribution of these haloes 
(Mo \& White 1996; Catelan et al. 1997; Sheth \& Lemson 1999).  
In these models, the haloes are biased tracers of the 
underlying dark matter distribution.  In general, this bias 
depends on the halo mass, and, for a given mass range, it is 
a nonlinear and stochastic function of the underlying dark 
matter density field.  

The shape of the mass function (and its evolution) predicted 
by these models is in reasonable agreement with what is 
measured in numerical simulations of hierarchical clustering 
from Gaussian initial conditions (e.g. Lacey \& Cole 1994).  
Although this agreement is by no means perfect (see 
Fig.~\ref{gifnm} below), to date, the emphasis has been on 
how well the models fit the simulations (but see Tormen 1998).  
On the other hand, recent work has shown that, while the 
model predictions for the bias relation are in reasonable 
agreement with what is measured in numerical simulations for 
massive haloes, less massive haloes are more strongly clustered, 
or less anti-biased, than the models predict 
(Jing 1998; Sheth \& Lemson 1999; Porciani, Catelan \& Lacey 1998).  

In this paper, following a suggestion by Sheth \& Lemson (1999), we 
argue that this discrepancy between the bias model and simulation 
results arises primarily because the model mass functions are 
different from those in the simulations.  In Section~\ref{model} 
we provide a simple relation between the mass function and the 
large scale bias factor.  We then use the mass function measured 
in the simulations to show that this relation is accurate.  
Section~\ref{sims} shows that our model provides a reaonably good 
fit to the bias relation of less massive haloes as well as massive 
ones.  

\section{Mass functions and large scale biasing}\label{model}
Let $f(m,\delta)\,{\rm d}m$, where $\delta=\delta(z)$ is a 
function of redshift $z$, denote the fraction of mass that is 
contained in collapsed haloes that at $z$ have mass in the 
range ${\rm d}m$ about $m$.  
The associated unconditional mass function is 
\begin{equation}
n(m,\delta)\,{\rm d}m = {\bar\rho\over m}\,f(m,\delta)\,{\rm d}m
\end{equation}
where $\bar\rho$ is the background density.  
Let $f(m_1,\delta_1|M_0,\delta_0)$ denote the fraction of the 
mass of a halo $M_0$ at $z_0$ that was in subhaloes of mass 
$m_1$ at $z_1$, where $z_1>z_0$.  
The associated conditional mass function is 
\begin{equation}
{\cal N}(1|0) = {M_0\over m_1}\,f(m_1,\delta_1|M_0,\delta_0).  
\end{equation}
Finally, let $\bar N(m_1,\delta_1|M,V,z_0)$ denote the average 
number of $(m_1,\delta_1)$ haloes that are within cells of size 
$V$ that contain mass $M$ at $z_0$.  
The halo bias relation is defined by
\begin{equation}
\delta_{\rm h}(1|0) \equiv 
{\bar N(m_1,\delta_1|M,V,z_0)\over n(m_1,\delta_1)\,V} - 1.
\label{bias}
\end{equation}
Since $M/V\equiv \bar\rho (1+\delta)$, this expression provides 
a relation between the overdensity of the halo distribution and 
that of the matter distribution.  
To compute this quantity, we need models for the numerator and 
the denominator.  This can be done as follows.  

An $M_0$ halo which collapsed at $z_0$ initially occupied a 
certain volume $V_0$ in the initial Lagrangian space.  
Let $\delta_0$ denote the initial overdensity within this $V_0$.  
If $|\delta_0|\ll 1$, then 
$M_0=\bar\rho\,(1+\delta_0)\,V_0\approx \bar\rho\,V_0$.  
So the mean Lagrangian space bias between haloes and mass 
is given by setting $\bar N(m_1,\delta_1|M_0,V,z_0) = {\cal N}(1|0)$ 
and $V=V_0$ in equation~(\ref{bias}) above.  Thus, 
\begin{equation}
\delta_{\rm h}^{\rm L}(1|0) = 
{{\cal N}(m_1,\delta_1|M_0,z_0)\over n(m_1,\delta_1)\,V_0} - 1 
.  
\label{biasl}
\end{equation}
This expression was first derived by Mo \& White (1996).
It can be expanded formally as a series in $\delta_0$:  
\begin{equation}
\delta_{\rm h}^{\rm L}(1|0) = 
\sum_{k} b_k^{\rm L}(m_1,\delta_1)\,\delta_0^k ,
\label{blseries}
\end{equation}
in which form it will appear later.  

To compute the bias relation in the evolved Eulerian space 
they suggested the following procedure.  Continue to assume 
that $\bar N(m_1,\delta_1|M_0,V,z_0)$ can be written as some 
${\cal N}(1|0)$, but now provide a new relation between 
the Eulerian variables $M_0$, $V$, and $z_0$, and the 
Lagrangian variables $M_0$, $V_0$ and $\delta_0$.  For any 
given $z_0$, Mo \& White used the spherical collapse model 
to write $\delta_0(M_0/\bar\rho V)$.  Recall that 
$M_0/\bar\rho V\equiv (1+\delta)$.  Thus, $\delta_0$ is a 
function of $\delta$, and, in general $\delta_0(\delta)$ 
is nonlinear.  However, when $|\delta|\ll 1$, then 
$\delta_0\approx\delta$.  
On large scales in Eulerian space, the rms fluctuation about 
the mean density is small.  Therefore, on sufficiently large 
scales, $|\delta|\ll 1$ almost surely.  
So, on large scales in Eulerian space, 
\begin{eqnarray}
\delta_{\rm h}^{\rm E}(1|0) &\equiv& 
{\bar N(m_1,\delta_1|M,V,z_0)\over n(m_1,\delta_1)\,V} - 1 \nonumber \\
&\approx& (1+\delta) \left[1 + \delta_{\rm h}^{\rm L}(1|0)\right] - 1
\nonumber \\
&\approx& \bigl[1 + b_{\rm Lag}(m_1,\delta_1)\bigr]\,\delta ,
\label{dhaloe}
\end{eqnarray}
where the final line follows from using equation~(\ref{blseries}), 
setting $\delta_0\approx\delta$, and so expanding to lowest order 
in $\delta$ (since $|\delta|\ll 1$), and writing $b_{\rm Lag}$ 
instead of $b_1^{\rm L}$.  This provides a simple linear relation 
between $\delta_{\rm h}^{\rm E}$ and $\delta$, where the 
constant of proportionality is 
\begin{equation}
b_{\rm Eul}(m_1,\delta_1) \equiv 1 + b_{\rm Lag}(m_1,\delta_1).  
\label{biase}
\end{equation}

Furthermore, Mo \& White (1996) argued that one can write 
\begin{equation}
\Bigl\langle \bigl(\delta_{\rm h}^{\rm E}\bigr)^2 \Bigr\rangle_{\rm V} 
\approx b^2_{\rm Eul}(m_1,\delta_1)\,
\Bigl\langle\delta^2\Bigr\rangle_{\rm V} ,
\end{equation}
where the average is over cells of size $V$ placed randomly in 
Eulerian space.  This expression neglects the effects of 
stochasticity on the bias relation (see Sheth \& Lemson 1999 
for details).  The left hand side is the volume average (over cells 
of size $V$) of the halo-halo correlation function, whereas 
$\langle\delta^2\rangle$ is the volume averaged correlation 
function of the dark matter.  Thus, in this limit, the ratio of 
the volume averaged correlation functions gives the square of the 
Eulerian bias factor.  

Now, $b_{\rm Eul}(m_1,\delta_1)$ is a function of halo mass only; 
it does not depend on scale.  So, in this large scale limit, the 
ratio of the volume averaged correlation functions is scale 
independent.  Therefore, the ratios of the correlation functions 
themselves should also be $b^2_{\rm Eul}$, independent of scale.  
However, correlation functions and their volume integrals are just 
differently weighted integrals over the associated power spectrum.  
If $b_{\rm Eul}$ is independent of scale on large scales, then the 
ratio of the halo and dark matter power spectra should also be 
$b^2_{\rm Eul}$, independent of wavenumber $k$ for small wavenumbers.  
Thus, in the large scale limit, we expect the bias factor to be 
the same, regardless of whether it was computed using the 
counts-in-cells method, the correlation functions themselves, 
or power spectra.  Moreover, in this limit, the Eulerian bias is 
trivial to compute if the Lagrangian space bias is known.  

The Lagrangian bias factor is different for haloes of different 
masses.  The work of Mo \& White (1996) shows that the exact form 
of this dependence can be computed provided that both the 
conditional and the unconditional mass functions are known 
(equation~\ref{biasl}).  Our contribution is to add one more 
simple step to the argument.  Namely, on the large scales 
discussed above, the peak background split should be a good 
approximation (e.g. Bardeen et al. 1986; Cole \& Kaiser 1989).  
In this limit, when expressed in appropriate variables, the 
conditional mass function is easily related to the unconditional 
mass function.  Therefore, on the large scales where the peak 
background split should apply, the bias relation can be 
computed even if only the unconditional mass function is known.   

To summarize: we have used the peak background split to argue that the 
dependence of $b_{\rm Lag}$ (and so $b_{\rm Eul}$ also) on halo mass 
should be sensitive to the shape of the unconditional mass function; 
different models for the unconditional mass function will produce 
different bias relations.  In particular, if the unconditional mass 
function is different from the one in simulations, the predicted bias 
will also be different.  In the next section we will use the 
unconditional mass function measured in simulations as input to the 
formulae above to compute the associated halo to mass bias relation.  

\section{The simulations}\label{sims}
In what follows, we will study the halo distributions which 
formed in what are known as the GIF simulations, which have been
kindly made available by the GIF/Virgo collaboration (e.g. Kauffmann 
et al. 1998a). We will show results for three choices of the initial 
fluctuation distribution belonging to the cold dark matter family: 
a standard model 
(SCDM: $\Omega_0 = 1$, $\Omega_\Lambda = 0$, 
$h=0.5$\footnote{The Hubble constant is $H_0=100h\ {\rm km\ s}^{-1}{\rm
Mpc}^{-1}$}), 
an open model 
(OCDM: $\Omega_0 = 0.3$, $\Omega_\Lambda = 0$, $h=0.7$) 
and a flat  model with non-zero cosmological constant 
($\Lambda$CDM: $\Omega_0 = 0.3$, $\Omega_\Lambda = 0.7$, $h=0.7$).  
These simulations were performed with $256^3$ particles each, in 
a box of size $L = 85\ {\mathrm Mpc}/h$ for the SCDM run, 
and $L = 141\ {\mathrm Mpc}/h$ for the OCDM and $\Lambda$CDM runs. 
The models and resolution are similar to those presented by 
Jing (1998).  

We measure mass functions in the simulations in the usual way, 
using a spherical overdensity group finder (see Tormen 1998 for 
details).  Previous authors have computed the large scale bias 
factor by using the ratio of the volume averaged halo and mass 
correlation functions, obtained using a counts-in-cells procedure 
(Mo \& White 1996; Sheth \& Lemson 1999).  Others have used the 
ratio of the correlation functions directly 
(Jing 1998; Porciani, Catelan \& Lacey 1998).  
Here, we will compute the bias factor by measuring the ratio of 
the power spectra of the haloes and the dark matter.  
On the large scales on which the peak background split should 
apply, all these measures should be equivalent.  Indeed, the extent 
to which these procedures all agree is a measure of the linearity 
of the large scale bias relation (equation~\ref{dhaloe}), and the 
accuracy of the assumption that the effects of stochasticity on 
the bias relation are trivial to estimate.  

\begin{figure}
\centering
\epsfxsize=\hsize\epsffile{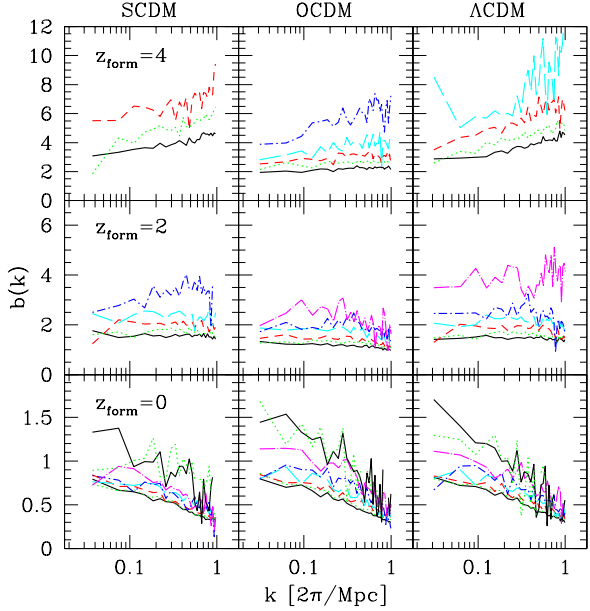}
\caption{The square root $b(k)$ of the ratio of the power spectra of 
dark matter haloes to that of the dark matter, computed at the time 
the haloes were identified:  $z=0$, $z=2$ and $z=4$ for the top, middle
and bottom rows, respectively.  The different line types in each panel 
correspond to haloes of different mass ranges (mass bins increase 
in size by factors of two).  The more massive haloes at a given 
redshift typically have larger values of $b(k)$.}
\label{gifbk}
\end{figure}
We will begin our comparison with simulations by showing the 
ratios of the halo and dark matter power spectra.  
The various panels in Fig.~\ref{gifbk} show the square root of 
$b^2(k)\equiv P_{\rm halo}(k)/P_{\rm matter}(k)$ for haloes 
identified at $z_{\rm form}=0$, $2$, and $4$, at which time both 
$P_{\rm halo}(k)$ and $P_{\rm matter}(k)$ were computed.  
The various line types in each panel correspond to haloes with 
mass in different mass ranges.  The lowest mass halo we consider 
has twenty particles, and the mass bins increase in size by factors 
of two.  Thus, the lower limit of the $m$th bin is at 
$10 \times 2^m$ particles per halo ($m>0$).  
One consequence of this is that the highest redshift outputs of 
the GIF simulations are only able to probe high values of $b(k)$, 
whereas lower redshift outputs primarily probe lower values of 
$b(k)$.  

An optimist would argue that the figure shows that $b(k)$ is 
approximately independent of $k$ for small $k$, though this 
approximation is more accurate when $b(k)\sim 1$ than otherwise.  
At a given redshift, the small $k$ value of $b(k)$ increases 
with halo mass.  
If the model presented in the previous section 
is correct, then we should be able to use the shape of the halo 
mass function to model this dependence accurately.  
We turn, therefore, to the shape of the halo mass function in 
the GIF simulations.  

The shape of the unconditional mass function is expected to 
depend on the initial fluctuation distribution 
(e.g. Press \& Schechter 1974).  If the initial 
distribution is Gaussian with a scale free spectrum, then 
the unconditional mass function can be expressed in units 
in which it has a universal form that is independent of 
redshift and power spectrum.  Namely, let 
$\nu\equiv [\delta_{\rm c}(z)/\sigma(m)]^2$, where 
$\delta_{\rm c}(z)$ is that critical value of the initial 
overdensity which is required for collapse at $z$, computed 
using the spherical collapse model, and $\sigma(m)$ is the value 
of the rms fluctuation in spheres which on average contain mass 
$m$ at the initial time, extrapolated using linear theory to $z$.  
For example, in models with $\Omega_0=1$ and $\Lambda_0=0$, 
$\delta_{\rm c}(z)=1.68647$, and $\sigma(m)\propto (1+z)^{-1}$.  
Then, for initially scale free spectra in models with 
$\Omega_0=1$ and $\Lambda_0=0$, 
\begin{equation}
\nu\,f(\nu)\equiv m^2\,{n(m,z)\over\bar\rho}\,
{{\rm d}\log m\over {\rm d}\log\nu} 
\label{nufnu}
\end{equation}
has a universal shape (Press--Schechter 1974; Peebles 1980; 
Bond et al. 1991).  
It is not obvious that this scaling holds for more general 
initial power spectra, such as those of the GIF simulations.  

\begin{figure}
\centering
\epsfxsize=\hsize\epsffile{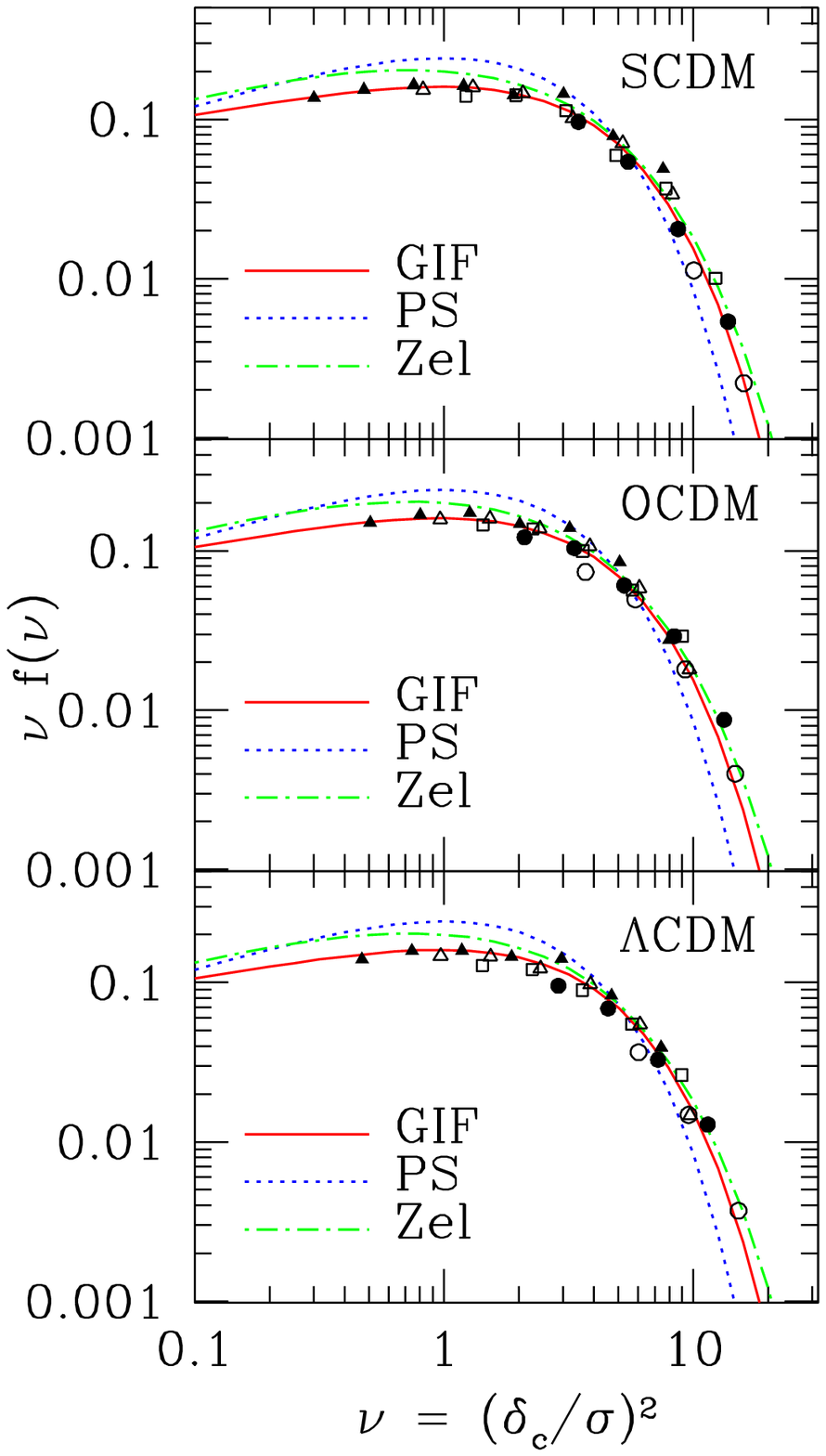}
\caption{The unconditional mass functions from five different 
output times (filled triangles, open triangles, open squares, 
filled circles, open circles show results for $z=0$, $z=0.5$, 
$z=1$, $z=2$ and $z=4$) in the GIF simulations plotted as a 
function of the scaled variable $\nu$.  
Dotted curve shows the Press--Schechter prediction, 
dot-dashed curve shows the mass function associated with the 
Zeldovich approximation, and solid curve shows our modified fitting 
function.}
\label{gifnm}
\end{figure}

The different symbols in Fig.~\ref{gifnm} show the unconditional 
mass function at different output times in the GIF simulations 
plotted as a function of the scaled variable $\nu$.  
The dotted line shows the Press--Schechter formula for 
$\nu\,f(\nu)$; note the discrepancy at both high and low $\nu$.  
The dot-dashed line, which fits the data slightly better than 
the dashed line, is the mass function computed using the Zeldovich 
approximation by Lee \& Shandarin (1998) (see Appendix A).  
The solid line through the data points shows a modification to 
the Press--Schechter function.  It has the form:
\begin{equation}
\nu\,f(\nu) = A\,\left(1 + {1\over \nu'^{p}}\right)\ 
\left({\nu'\over 2}\right)^{1/2}\,{{\rm e}^{-\nu'/2}\over\sqrt{\pi}} ,
\label{giffit}
\end{equation}
where $\nu'=a\,\nu$, with $a=0.707$ and 
$\nu=(\delta_{\rm c}/\sigma)^2$ as described in the previous 
paragraph, $p=0.3$, and $A$ is determined by requiring that the 
integral of $f(\nu)$ over all $\nu$ give unity.  
The original Press--Schechter formula has $a=1$, $p=0$, and 
$A=1/2$.  

Recall that, in principle, the shape of $\nu\,f(\nu)$ can depend on 
the shape of the initial power spectrum, and on redshift.  Since 
the different symbols in each panel overlap each other reasonably 
well, Fig.~\ref{gifnm} shows that, in the GIF simulations at least, 
it is a good approximation to say that the mass functions scale 
similarly to the scale free case.  So, although in principle we could 
have allowed $a(z,\Omega_0,\Lambda_0)$ and $p(z,\Omega_0,\Lambda_0)$ 
in equation~(\ref{giffit}) (in fact, we could even have allowed a 
different functional form for each output time of each model), 
Fig.~\ref{gifnm} shows that, in practice, equation~(\ref{giffit}) 
with the same value of $p$ provides a reasonably good description 
of the mass function at all output times.  This suggests that the 
dynamics of collapse is sensitive to $\nu$, and not to the mass 
scale itself.  That the same value of $p$ is a good approximation 
to $\nu\,f(\nu)$ in all three panels is presumably a consequence of 
the fact that the shapes of the initial power spectra in these 
models are not very different.  Therefore, in what follows, rather 
than finding best fit shapes to $\nu\,f(\nu)$ at each output time 
of each model, we will simply use this one function with the given 
constants $a$ and $p$.  

\begin{figure*}
\centering
\epsfxsize=\hsize\epsffile{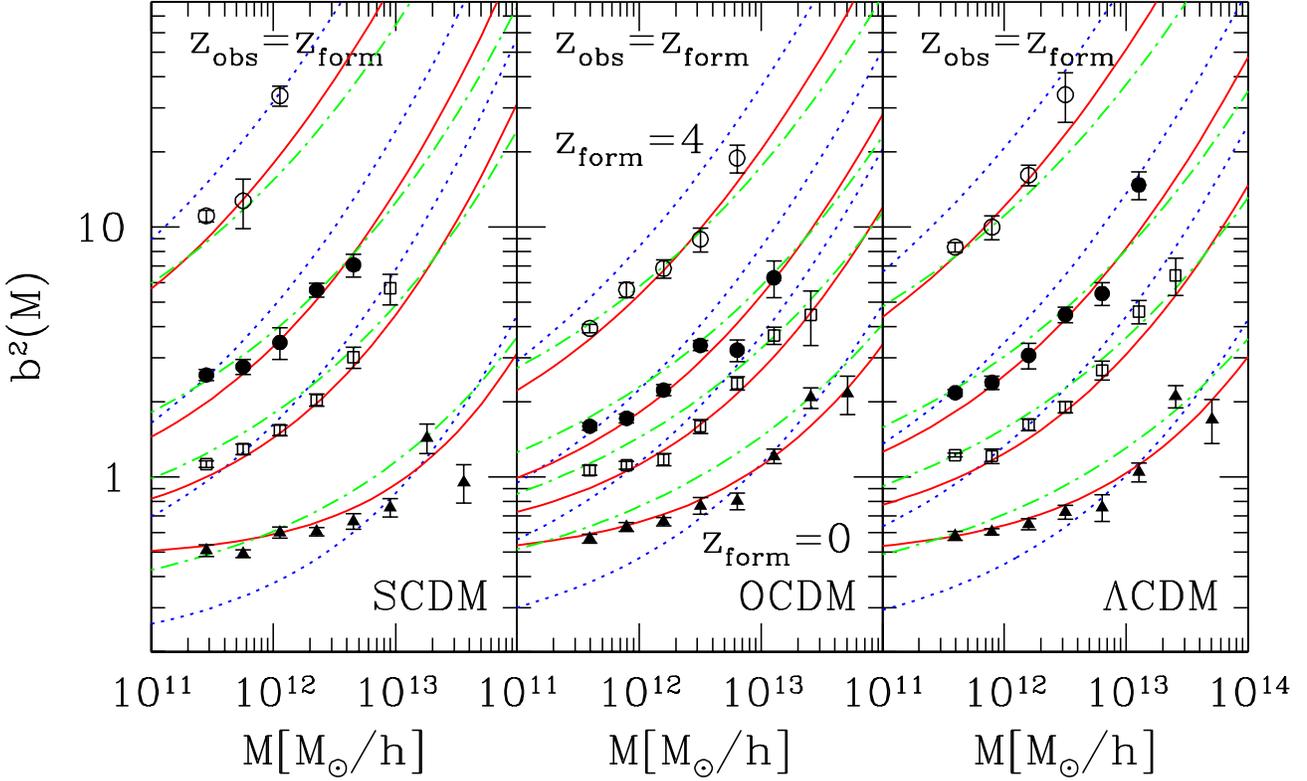}
\caption{The large scale Eulerian bias relation at 
$z_{\rm obs}=z_{\rm form}$ between haloes which are identified 
at $z_{\rm form}$, and the mass.  Solid curves show the relation 
we predict using the GIF mass function, dotted curves show 
the relations which follow from the Press--Schechter mass function,
and dot-dashed curves show the relations associated with the 
Zeldovich mass function.} 
\label{gifbias}
\end{figure*}

This fortuitous rescaling of the mass function is useful because 
it means that we need compute the peak-background split limit 
associated with the unconditional mass function only once, rather 
than having to compute it for each output time of each model.  
Simply set $\nu\to\nu_{10}$ in the expression above, where 
\begin{displaymath}
\nu_{10}\equiv 
{[\delta_{\rm c}(z_1)-\delta_{\rm c}(z_0)]^2\over 
\sigma^2(m_1)-\sigma^2(M_0)} \approx 
{(\delta_1-\delta_0)^2\over \sigma_1^2} \approx 
\nu_1 \left(1 - {2\delta_0\over \delta_1}\right),
\end{displaymath} 
with $\delta_1\equiv\delta_{\rm c}(\Omega_1)$, 
where $\Omega_1\equiv\Omega(z_1)$, $\sigma_1\equiv \sigma(m_1)$ 
scaled using linear theory to $z_1$ and 
$\nu_1\equiv (\delta_1/\sigma_1)^2$, and require that 
$f(m_1|M_0)\,{\rm d}m_1 = f(\nu_{10})\,{\rm d}\nu_{10}$.

Once the unconditional and conditional mass functions are 
known, the peak background split bias in Lagrangian space 
is easy to compute.  It is 
\begin{equation}
\delta^{\rm L}_{\rm h}(1|0) \approx 
\left[a\nu_1 - 1 + {2p\over 1 + (a\nu_1)^p}\right]\, 
{\delta_0\over\delta_1} = 
b^{\rm GIF}_{\rm Lag}(m_1,\delta_1)\,\delta_0,
\label{biaspbs}
\end{equation}
where the final expression defines $b_{\rm Lag}^{\rm GIF}$.  
When $a=1$ and $p=0$ the mass function has the Press--Schechter 
form, and this formula reduces to the one given by 
Cole \& Kaiser (1989) and Mo \& White (1996).  
However, the GIF mass function has $a=0.707$ and $p=0.3$.  
Since $\nu_1$ increases with increasing halo mass, the final 
term in the expression above is negligible for massive haloes.  
In this limit, the expression above resembles the original 
Press--Schechter based formula, except for the factor of 
$a$.  Since $a<1$, our formula predicts that massive haloes are 
slightly less biased than the original Press--Schechter based 
formula predicts.  For less massive haloes the extra term cannot 
be ignored.  It is positive for all $\nu_1$, so our formula 
predicts that less massive haloes are more positively biased 
(or less anti-biased) than the Mo \& White formula suggests.  
The bias in Eulerian space is got by substituting this 
expression into equation~(\ref{biase}):
\begin{equation}
b^{\rm GIF}_{\rm Eul}(m_1,\delta_1) = 1 + {a\nu_1 - 1\over \delta_1} + 
{2p/\delta_1\over 1 + (a\nu_1)^p} .
\label{begif}
\end{equation}  
Both the trends discussed above remain true in Eulerian space.  

\begin{figure}
\centering
\epsfxsize=\hsize\epsffile{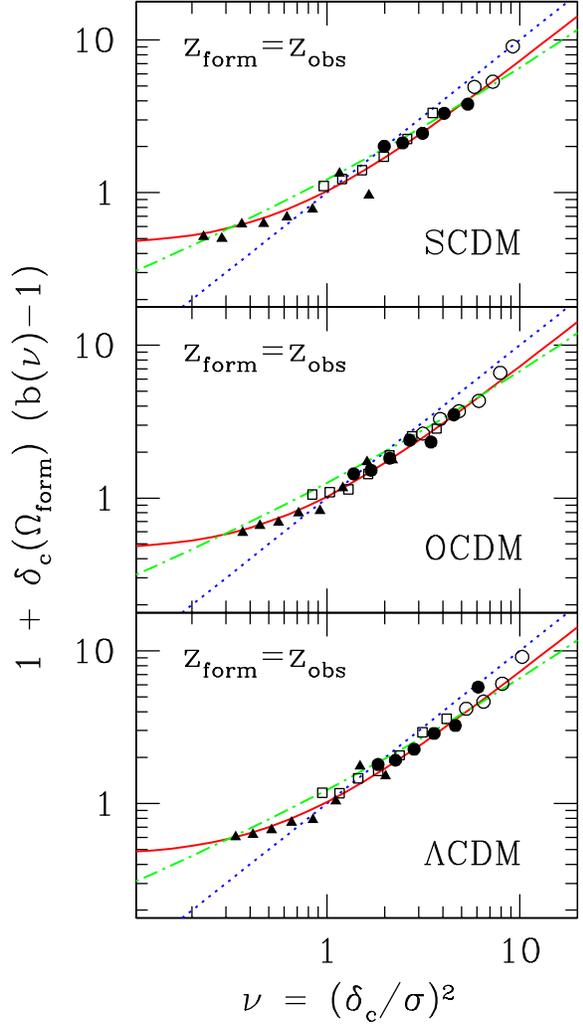}
\caption{The large scale Eulerian bias relation of the 
previous figure, rescaled as indicated by the axis labels.  
Solid curves show the relation we predict using the 
GIF mass function, dot-dashed curves show the corresponding 
relation predicted using the Zeldovich mass function, and 
dotted curves show the relation which follows from the 
Press--Schechter mass function.  } 
\label{nubias}
\end{figure}

The different symbols in Fig.~\ref{gifbias} show the large scale 
bias relation for haloes that are identified at $z_{\rm form}$ 
and are observed at $z_{\rm obs}=z_{\rm form}$.  That is, they 
show the small $k$ values of $b^2(k)$ obtained as the value 
predicted at $k=2\pi/L$ by a linear least-square fit to the 
12 left-most data points.  Error bars show the formal least-square
error.  In this way we tried to estimate the actual large-scale 
bias even when $b(k)$ on the largest scales probed by the simulations 
shows a dependence on $k$.  One might argue that the weak scale 
dependence of $b(k)$ in Fig.~\ref{gifbk} means that the symbols 
in Fig.~\ref{gifbias} probably slightly under/overestimate 
the true large scale (small $k$) value for small/large values 
of $b(k)$.  

Each panel shows four sets of curves, corresponding to haloes 
which formed at $z_{\rm for}=4$, $2$, $1$, and $0$.  
The three curves for each redshift show how the bias relation 
computed using our peak background split model depends on the 
unconditional mass function.  Solid and dotted curves show 
equation~(\ref{begif}) for the GIF ($a=0.707$ and $p=0.3$) and 
the Press--Schechter ($a=1$, $p=0$) mass functions, respectively.
The dot-dashed curves show the bias associated with the 
Zeldovich mass function (equation~\ref{bezel}).  
As with the mass function, the dot-dashed curves fit the data 
better than the dotted curves, and slightly worse than the 
solid ones.  The solid curves fit the data reasonably well for 
all masses and redshifts.  

There are small discrepancies between our GIF based predictions 
and the symbols at large and small $b(M)$.  Since they are in the 
sense discussed earlier, we suspect some of the discrepancy 
arises from the weak scale dependence of $b(k)$.  
Since these discrepancies are small, and since our measured values 
of $b(M)$ at small $M$ when $z=0$ are very similar to those in 
Fig.~3 of Jing (1998), it appears that using the ratio of the power 
spectra to define $b_{\rm Eul}$ gives nearly the same result as 
using the ratio of the correlation functions.  As discussed at the 
end of the previous section, this is expected if the bias is 
independent of scale.  

Suppose that the unconditional mass function $\nu\,f(\nu)$ 
really was independent of $z_{\rm form}$ (recall that our 
Fig.~\ref{gifnm} shows this is a good approximation).  
Then our peak background split argument says that, after 
appropriate transformations, it should be possible to present 
the large scale bias relation as function of $\nu$ only.  
Namely, equation~(\ref{begif}) suggests that we should be able 
to scale all the results for the different 
$z_{\rm form}$ above to one plot of the product 
($\delta_1\,b_{\rm Lag}$) versus $\nu$.  
Fig.~\ref{nubias} shows the result.  The symbols 
show the simulation data from the same output times as before.  
They were obtained by transforming the large scale Eulerian 
bias relation as indicated by the labels on the axes.  
The curves show the theoretical predictions for the three 
different mass functions we have been considering.  

There are two points to be made about this plot.  
The first is that our peak background split model (solid curves) 
provides a reasonable description of the results.  
The main reason for including this plot is to show that the 
scatter about the solid curves in our plots is approximately 
the same as the scatter about the mass function fits in 
Fig.~\ref{gifnm}.  As we argued above, we have no a priori 
reason to expect such a scaling if our relation between the 
unconditional mass function and the Eulerian bias were not 
accurate.  

Finally, we can also study the case in which haloes form at 
high redshift, but are observed at lower redshift:  
$z_{\rm form}>z_{\rm obs}$.  In this case, the large scale 
Eulerian bias should be described by our equations~(\ref{biase}) 
and~(\ref{begif}), but with 
\begin{displaymath}
\nu_1 \equiv 
{\delta^2_{\rm c}(\Omega_{\rm form})\over\sigma^2(m)}
\qquad{\rm and}\qquad 
\delta_1 \equiv {D(z_{\rm obs})\over D(z_{\rm form})}\,
\delta_{\rm c}(\Omega_{\rm form})
\end{displaymath}
where $D(z)$ is the linear theory growth factor, and $\sigma(m)$ 
is scaled using linear theory to its value at $z_{\rm form}$.   

\begin{figure}
\centering
\epsfxsize=\hsize\epsffile{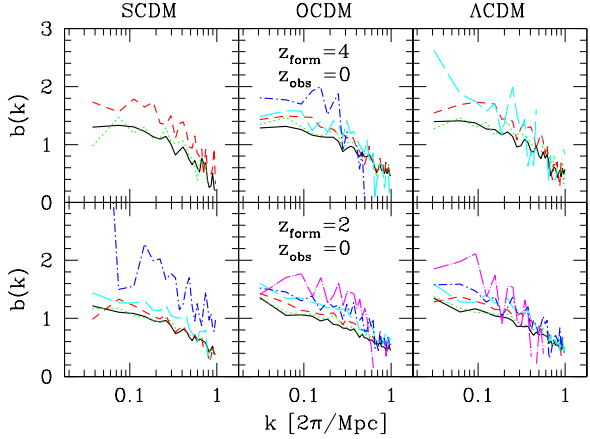}
\caption{The square root of the ratio of the halo and matter 
power spectra for haloes that formed at $z_{\rm form}=4$ (top 
panels) and $z_{\rm form}=2$ (bottom panels) but were observed 
at $z_{\rm obs}=0$.  }
\label{twozbks}
\end{figure}

\begin{figure}
\centering
\epsfxsize=\hsize\epsffile{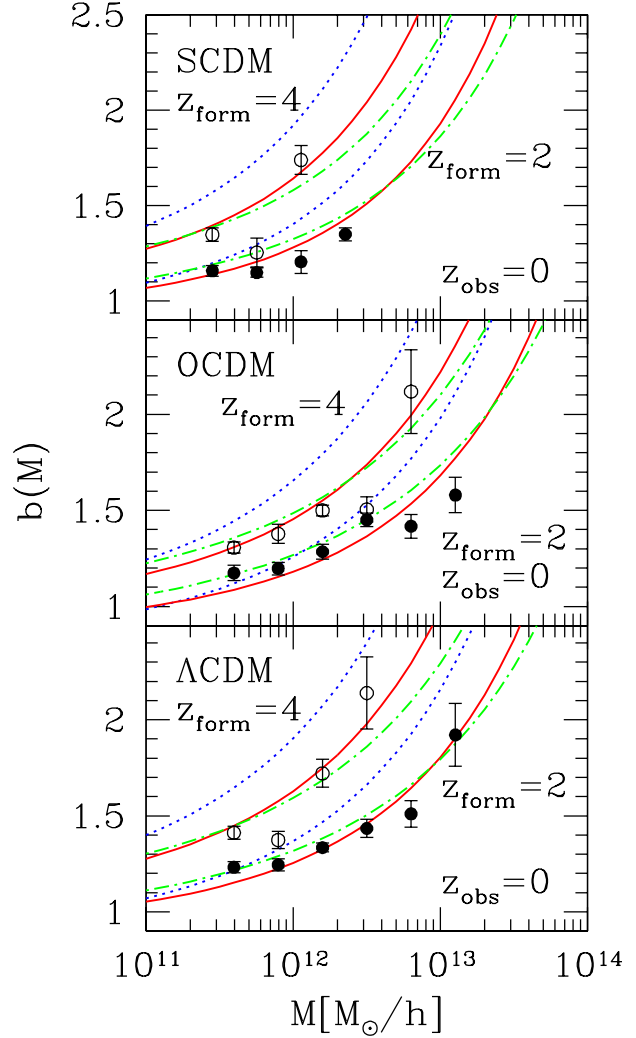}
\caption{The large scale Eulerian bias relation at 
$z_{\rm obs}=0$ for haloes that formed at $z_{\rm form}$.  
Solid curves show the relation we predict using the GIF mass 
function, dotted and dot-dashed curves show the relations which 
follow from the Press--Schechter and Zeldovich mass functions, 
respectively. }
\label{twozs}
\end{figure}

Fig.~\ref{twozbks} shows the ratio of the halo and matter power 
spectra for haloes which formed at $z_{\rm form}>0$ but were 
studied at $z_{\rm obs}=0$.  Thus, it is similar to 
Fig.~\ref{gifbk}, except that there $z_{\rm obs}=z_{\rm form}$.  
There are a few obvious differences:  some of the $b(k)$ curves 
are not particularly flat, even on the largest scales.  
Those curves that do asymptote nicely (at small $k$) do so at a 
value of $b(k)$ that is closer to unity than when 
$z_{\rm obs}=z_{\rm form}$.  Moreover, the scale dependence of 
$b(k)$ is different from before:  whereas previously $b(k)$ for 
these haloes increased with increasing $k$, now it decreases.  

Estimating the large-scale value of $b(k)$ by a least-square fit 
as done for Fig.~\ref{gifbias} yields the symbols shown in 
Fig.~\ref{twozs}.  
Most of the symbols are well fit by our simple formula for the 
large scale bias relation.  The symbols that are not well fit by 
our formula (typically those in the highest mass bins) almost 
always correspond to haloes in a mass range in which the power 
spectra were not particularly flat.  We conclude, therefore, 
that our formula for the bias relation is reasonably accurate 
in this two epoch context as well.  

\section{Discussion}
Collapsed haloes are biased tracers of the dark matter 
distribution.  This bias depends on halo mass.  
In general, to quantify this bias requires a detailed 
knowledge of the merger histories of dark matter haloes.  
We used the peak background split to argue that, on large 
scales, this detailed information is unnecessary:  
knowledge of only the shape of the unconditional mass function 
is sufficient to compute a good approximation to the large 
scale bias relation.  This fact was anticipated by 
Sheth \& Lemson (1999), but seems to have been unnoticed before.  

For example, our approach is different from, but consistent with, 
the recent work of Jing (1998) and Porciani, Catelan \& Lacey (1998). 
Following Sheth \& Lemson's demonstration that the bias in Lagrangian 
space was not well described by standard models, Porciani et al. 
quantified the discrepancy.  They showed that using 
equation~(\ref{biase}) to transform Jing's (1998) fitting formula 
for the Eulerian bias provides a reasonably good fit to the 
Lagrangian bias factor.  In other words, their results show that 
equation~(\ref{biase}) is relatively accurate.  They then provided 
fitting functions for the Lagrangian bias relation.  However, since 
we have argued that the Lagrangian bias is related to the halo mass 
function, we have chosen to provide a fit to the mass function 
instead.  

In so doing, we showed that a simple modification 
(equation~\ref{giffit}) of the Press--Schechter formula provides a 
good fit to the unconditional halo mass function in the SCDM, OCDM 
and $\Lambda$CDM models we studied (Fig.~\ref{gifnm}).  This 
allowed us to derive a single, simple, physically motivated 
formula for the large scale bias relation (equation~\ref{begif}) 
that can be used for haloes of all masses at all times in all 
three cosmologies.  Comparison of the predictions of our model with 
numerical simulations showed good agreement.  We showed that this 
was true for the case in which haloes are observed at the same time 
as when they are first identified ($z_{\rm obs}=z_{\rm form}$), 
and also for the case in which haloes form at $z_{\rm form}$, evolve, 
and are observed at some later time $z_{\rm obs}<z_{\rm form}$.  

If haloes are significantly more/less massive than the typical 
$M_*$ halo at the time of formation, then our formulae predict 
that they should be less/more biased than results based on the 
Press--Schechter formula would predict.  Since galaxies are 
thought to form in small, i.e., $M/M_*(z_{\rm form})<1$, haloes 
(Kauffmann et al. 1998b; Baugh et al. 1998), our results 
are useful for galaxy formation models.  
The first generation of stars and high redshift quasars are 
thought to be associated with haloes having 
$M/M_*(z_{\rm form})\gg 1$ (Haehnelt, Natarajan \& Rees 1998; 
Haiman \& Loeb 1998), so our results (for the number density as 
well as the bias factor) may also be useful for studies of the 
reionization history of our Universe.  

Although we showed how the bias relation depends on the 
unconditional mass function (for example, the bias relation 
associated with the simple Press--Schechter spherical collapse 
mass function is different from that associated with the 
Zeldovich mass function), we did not provide a derivation 
of the shape of the mass function itself.  Since the mass 
function in the GIF simulations is different from those 
predicted by simple applications of the spherical collapse 
model or by the Zeldovich approximation, our model is still 
far from complete.  We hope that our results will stimulate 
work aimed at rectifying this.  

In this regard, we think it important to stress the following 
fact.  Bond et al. (1991) used a random walk barrier crossing 
model to derive the Press--Schechter unconditional mass function 
from the statistics of the initial fluctuation field.  
In their model, the Press--Schechter mass function is associated 
with the first crossing distribution of a barrier of constant 
height.  The barrier height is constant because, in the spherical 
collapse model, there is a critical overdensity $\delta_{\rm c}$ 
required for collapse, and this value is independent of the mass 
of the collapsed object.  If this same barrier crossing model is 
to yield the GIF mass function, then the barrier shape must 
depend on mass.  We have solved for this barrier shape, and are 
in the process of developing and testing the detailed predictions 
of such a moving barrier model further.  

\section*{ACKNOWLEDGMENTS}

GT acknowledges financial support from a postdoctoral fellowship 
at the Astronomy Department of the University of Padova.
We thank Simon White and the Virgo consortium for providing us 
access to the GIF simulations.

\appendix
\section{Large scale bias and the Zeldovich approximation}
The universal unconditional mass function associated with the 
Zeldovich approximation is given by equation~(\ref{nufnu}) 
with 
\begin{displaymath}
\nu f(\nu) = Z_{\rm a}(\nu') + Z_{\rm b}(\nu') + Z_{\rm c}(\nu'), 
\end{displaymath}
where 
\begin{eqnarray*}
Z_{\rm a}(\nu')\!\!\! &=&\!\!\! {25\over 4} \sqrt{10\nu'\over\pi}
\left({5\nu'\over 3}-{1\over 12}\right)
\exp\left(-5\nu'\over 2\right)
{\rm erfc}\left(\sqrt{2\nu'}\right) \nonumber \\
Z_{\rm b}(\nu') \!\!\! &=& \!\!\! {25\over 16}\sqrt{15\nu'\over\pi} \ 
\exp\left(-15\nu'\over 4\right)\,
{\rm erfc}\left(\sqrt{3\nu'\over 4}\right) \nonumber \\
Z_{\rm c}(\nu') \!\!\! &=&\!\!\!  -{125\over 12}{\sqrt{5}\over\pi}\ 
\nu'\,\exp\left({-9\nu'\over 2}\right),
\end{eqnarray*}
and 
\begin{equation}
\nu' = \left(\lambda_{\rm c}/\delta_{\rm c}\right)^2\,\nu = 
\left(\lambda_{\rm c}/\delta_{\rm c}\right)^2\,(\delta_{\rm c}/\sigma)^2
\end{equation}
(Lee \& Shandarin 1998).  In these variables, the mass function 
has two free parameters:  the usual spherical collapse 
$\delta_{\rm c}$, and the ratio $(\lambda_{\rm c}/\delta_{\rm c})$.  
The dot--dashed curve in Fig.~\ref{gifnm} shows this function with 
the value used by Lee \& Shandarin:  $\lambda_{\rm c}=0.38$.  

One way to approximate the associated conditional mass function is 
to use the peak background split discussed in the main text.  
In this approximation, one simply sets 
\begin{displaymath}
\nu' = 
{(\lambda_{\rm c1}-\lambda_{\rm c0})^2\over \sigma_1^2-\sigma_0^2} 
\approx \nu'_1\,(1 - \delta_{\rm c0}/\delta_{\rm c1})^2, 
\end{displaymath}
into the expression for $f(\nu)\,{\rm d}\nu$ above.  
Although we have not done so here, in principle, one could also 
allow $\lambda_{\rm c}(z,\Omega_0,\Lambda_0)$.  
In our case, the associated Lagrangian large scale bias is 
given by equation~(\ref{biasl}), so 
\begin{eqnarray}
\delta_{\rm h}^{\rm L}(1|0) &\approx& {\delta_0\over \delta_1}\ 
\Biggl[ 5\,\nu'_1 - 1 - 2\,Z_{\rm a}(\nu'_1)\,
{20\nu'_1\over 20\nu'_1 - 1}
\nonumber \\
&& \qquad\qquad\qquad + {5\over 4}\,
\Bigl(2\nu'_1\,Z_{\rm b}(\nu'_1) - Z_{\rm c}(\nu'_1)\Bigr) \Biggr] 
\nonumber\\
&\equiv& b^{\rm Zel}_{\rm Lag}(m_1,\delta_1)\,\delta_0 .
\label{biaszel}
\end{eqnarray}
The Eulerian space peak background split bias factor for the 
Zeldovich approximation is given by inserting this 
expression for $b_{\rm Lag}$ in equation~(\ref{biase}):
\begin{equation}
b^{\rm Zel}_{\rm Eul}(m_1,\delta_1) = 
1 + b^{\rm Zel}_{\rm Lag}(m_1,\delta_1).
\label{bezel}
\end{equation}

\end{document}